# A Peer-to-Peer Energy Trading Framework in Distribution Systems Considering Network Constraints

Saeed Nasiri

*Abstract*—With the widespread adoption of Renewable Energy Sources (RESs) in low-voltage distribution systems, opportunities for energy trading among peers have emerged. In particular, the advent of distributed ledgers and blockchain technologies has catalyzed the application of Peer-to-Peer (P2P) economic concepts in decentralized, small-scale energy trading. This paper focuses on the critical physical layer aspects of transactions within the context of P2P energy trading, with a specific emphasis on addressing network constraints. Key challenges include maintaining margins for over/under voltage, voltage balance, and preventing congestion, all of which must be upheld during P2P energy exchanges. To address these challenges, we propose a novel analytical approach tailored to distribution networks. Furthermore, we introduce the Block Double Auction (BDA) mechanism as the P2P market mechanism for determining the acceptance or rejection of P2P transactions. The effectiveness of our proposed method is validated using the IEEE 33-node distribution test system, demonstrating its robust capabilities.

*Index Terms*—Low voltage systems, P2P markets, Rooftop PV.

## NOMENCLATURE

| | |
|---|---|
| $p_i$ | Active power injection at node i |
| $Q_i$ | Reactive power injection at node i |
| $|v_i|$ | Voltage magnitude at node i |
| $\delta_i$ | Voltage angle at node i |
| $|Y_{ij}|$ | Magnitude of admittance matrix entry corresponding to nodes i and j |
| $\theta_{ij}$ | Angle of admittance matrix entry corresponding to nodes i and j |
| $\Delta P$ | Active power injection variation |
| $\Delta Q$ | Reactive power injection variation |
| $\Delta|V|$ | Voltage magnitude variation |
| $\Delta\delta$ | Voltage angle variation |
| $J_{PV}$ | Jacobian matrix block corresponding to voltage magnitude and active power injections |
| $J_{QV}$ | Jacobian matrix block corresponding to voltage magnitude and reactive power injections |
| $J_{P\delta}$ | Jacobian matrix block corresponding to voltage angle and active power injections |
| $J_{Q\delta}$ | Jacobian matrix block corresponding to voltage angle and reactive power injections |
| $VUF$ | Voltage unbalance factor |
| $V^1$ | Positive sequence voltage |
| $V^2$ | Negative sequence voltage |
| $V^0$ | Zero sequence voltage |
| $\alpha$ | Maximum voltage unbalance level |
| $a, b, c$ | Phases |
| $\Delta I$ | Current variation |
| $z_{ij}$ | Line impedance between nodes i and j |
| $a^1$ | $1\angle 120°$ |
| $a^2$ | $1\angle 240°$ |
| $\| \cdot \|_p$ | Norm L-p |
| $\Delta I_{ij}^{max}$ | Maximum allowable Current variations |
| $la, lb, lc$ | Lines sets belonging to each phase |

## I. Introduction

THE increasing adoption of small-scale Distributed Energy Resources (DERs) by end-users within distribution systems has propelled both research endeavors and the industry to delve into Peer-to-Peer (P2P) electricity markets. This transition reflects a shift from the traditional role of consumers to that of prosumers, who actively engage in both electricity consumption and generation. Multiple drivers underpin this transformative trend. Notable among these drivers include the local supply of demand, incentivizing customers to invest in their own electricity generation, deferring the need for extensive transmission and generation infrastructure expansions, and operating the network with reduced external costs [1]. Moreover, the advent of blockchain technology presents promising opportunities for managing a multitude of participants within a market, obviating the necessity for a central intermediary [2]. This is particularly advantageous in scenarios where market mechanisms and settlement processes may prove cost-prohibitive and impractical due to the involvement of a large number of participants engaged in small-scale transactions [3-6].

While the precise requirements and roles of participants in Peer-to-Peer (P2P) markets remain open questions, these markets typically comprise four distinct layers [7]. The first layer, the business layer, encompasses a wide array of components, including the market mechanism, cash flow considerations, cost-benefit analyses for service providers, roles of Distribution System Operators (DSOs), coordination with local and wholesale markets, allocation of external costs, energy pricing strategies, and the management of byproducts markets [8, 9]. The second layer, the control layer, assumes responsibility for various control functions such as power flow control, voltage regulation, frequency stabilization, and reactive power management. Within this layer, a notable concept involves the development of control strategies that facilitate interaction between smart homes and appliances, enabling them to respond to price signals from the network. This concept is commonly referred to as Transactive Control [10-12].

The third layer encompasses the information and communication layer, which encompasses a diverse range of applications. These include end-user software designed for participation in the market, the implementation of robust cyber security measures, and the integration of sensors and smart meters. This layer plays a pivotal role in facilitating secure and seamless information exchange among peers and the Distribution System Operator (DSO) [13].

Last but certainly not least, the physical electrical network assumes the crucial role of delivering electricity traded among peers. Within this role, several challenging issues emerge. These encompass the design of protection systems and ensuring the reliability of the network, particularly in scenarios involving bidirectional current flows through the network and the incorporation of small-scale Distributed Generators (DGs). Many of these DGs are characterized by their stochastic and non-dispatchable nature as they endeavor to meet the energy demand. Additionally, it is worth emphasizing that all layers within the framework must exhibit seamless interoperability and secure coordination. Furthermore, the acceptance or rejection of an energy transaction should be predicated on technical constraints within the network, underscoring the paramount importance of maintaining network integrity.

The literature offers a range of papers, each predominantly focused on a specific layer within the P2P energy trading framework. In [14], a consortium blockchain-based approach is introduced for facilitating local P2P electricity trading among Plug-in Hybrid Electric Vehicles (PHEVs). [5] utilizes blockchain technology in conjunction with multi-agent modeling to develop a comprehensive model for P2P electricity trading. [15] puts forth a loss allocation method based on graph theory, particularly tailored for unbalanced distribution networks within transactive markets. In [16], a consensus-driven approach is proposed for P2P electricity markets, leveraging a multi-bilateral economic dispatch mechanism as the core market mechanism. [17] introduces a loss allocation method within blockchain-based P2P energy markets, accounting for transaction fees based on the network losses incurred by each transaction. [18] introduces the innovative concept of federated power plants, which can be established by appropriately incentivizing P2P energy exchanges. In [19], a bilateral contract network mechanism is presented as a means for facilitating P2P energy trading. [20] contributes by proposing a method to evaluate and accept or reject energy transactions based on the physical constraints inherent to the network.

Evidently, the majority of papers in the existing literature primarily focus on the business layer, specifically addressing market design and loss allocation. However, these approaches tend to fall short in the absence of due consideration for the physical layer's limitations. The method proposed in [20] faces several noteworthy challenges. Foremost among these challenges is its evaluation of each transaction in isolation, implying that each transaction, from one node to another, manifests itself physically within the network. It is evident that this assumption is primarily introduced to justify the approach for allocating transaction costs. Nevertheless, this assumption raises crucial issues. For instance, what transpires when two nodes from different phases engage in trade? Such interactions have the potential to induce voltage imbalances at network nodes, a condition that must remain within specified limits. Furthermore, adopting such an approach can lead to a reduction in social welfare since certain transactions may be erroneously accepted or rejected when evaluated independently. Moreover, the methodology overlooks voltage unbalance, a prominent challenge in low-voltage distribution networks.

This paper introduces an innovative analytical framework designed to holistically evaluate all Peer-to-Peer (P2P) transactions negotiated among peers. The outcome of this approach yields two distinct sets: one comprising acceptable transactions and another containing prone to be rejected ones that provides insight into the quantity of energy that pairs of peers are eligible to trade. The core criteria for acceptance within the transaction block encompass considerations related to over/under voltage, line congestion, and voltage unbalance. Transactions must successfully navigate these criteria to achieve acceptance.

The remainder of the paper is organized as follows. The proposed methodology for modeling voltage variations, voltage unbalance and congestion to assess the P2P transactions are described in section II. The proposed method for evaluating the P2P transactions block is presented in section III. The proposed market mechanism for P2P framework is discussed in section IV. The numerical study and discussion of efficacy of the proposed method are provided in section V. Finally, section VI concludes the paper.

## II. PROPOSED METHODOLOGY

This section is dedicated to an in-depth discussion of the proposed methodology for evaluating the feasibility of P2P transactions concerning the constraints within the low-voltage distribution network. As previously highlighted, three primary restrictive indices impose limitations on the volume of single-phase energy trading within distribution networks. Consequently, the acceptance or rejection of transactions is intricately tied to the margins associated with voltage violation, line congestion, and the degree of voltage unbalance. The ensuing subsections provide a comprehensive overview of the methodology employed for each of these crucial considerations.

### A. Over/under voltage

P2P transactions have the potential to infringe upon voltage limitations at various nodes within the network, contingent upon the traded energy volume and the locations of the participating peers. Consequently, the Distribution System Operator (DSO) must implement regulatory measures to effectively manage P2P trades.

In order to address voltage profile issues, we have developed a methodology based on sensitivity coefficients. It is well-established that the voltage at each node can be correlated with active and reactive power injections at different nodes, as described in equations (1a)-(1h). This correlation enables the assessment of the impact of power injections, whether they involve power absorption or injection, at various nodes within the network. These sensitivity coefficients essentially represent elements of the Jacobian matrix and must be determined using the base case of the system. The base case can correspond to either the most recent operational state of the network before the time slot during which new transactions are to be conducted.

$$\frac{\partial p_i}{\partial |v_i|} = 2|v_i||Y_{ii}|\cos(\theta_{ii}) + \sum_{j \neq i}|v_j||Y_{ij}|\cos(\theta_{ij} - \delta_i + \delta_j) \tag{1a}$$

$$\frac{\partial p_i}{\partial |v_j|} = |v_i||Y_{ij}|\cos(\theta_{ij} - \delta_i + \delta_j) \quad j \neq i \tag{1b}$$

$$\frac{\partial p_i}{\partial \delta_i} = \sum_{j \neq i}|v_i||v_j||Y_{ij}|\sin(\theta_{ij} - \delta_i + \delta_j) \tag{1c}$$

$$\frac{\partial p_i}{\partial \delta_j} = -|v_i||v_j||Y_{ij}|\sin(\theta_{ij} - \delta_i + \delta_j) \quad j \neq i \tag{1d}$$

$$\frac{\partial Q_i}{\partial |v_i|} = -2|v_i||Y_{ii}|\sin(\theta_{ii}) - \sum_{j \neq i}|v_j||Y_{ij}|\sin(\theta_{ij} - \delta_i + \delta_j) \tag{1e}$$

$$\frac{\partial Q_i}{\partial |v_j|} = -|v_i||Y_{ij}|\sin(\theta_{ij} - \delta_i + \delta_j) \quad j \neq i \tag{1f}$$

$$\frac{\partial Q_i}{\partial \delta_i} = \sum_{j \neq i}|v_i||v_j||Y_{ij}|\cos(\theta_{ij} - \delta_i + \delta_j) \tag{1g}$$

$$\frac{\partial Q_i}{\partial \delta_j} = -|v_i||v_j||Y_{ij}|\cos(\theta_{ij} - \delta_i + \delta_j) \quad j \neq i \tag{1h}$$

After determination of sensitivity coefficients, the variations of active and reactive powers can be related to variations of voltage magnitudes and angles using (2a) and (2b).

$$J = \begin{pmatrix} \dfrac{\partial P}{\partial |V|} & \dfrac{\partial P}{\partial \delta} \\ \dfrac{\partial Q}{\partial |V|} & \dfrac{\partial Q}{\partial \delta} \end{pmatrix} = \begin{pmatrix} J_{PV} & J_{P\delta} \\ J_{QV} & J_{Q\delta} \end{pmatrix} \tag{2a}$$

$$\begin{pmatrix} \Delta P \\ \Delta Q \end{pmatrix} = \begin{pmatrix} J_{PV} & J_{P\delta} \\ J_{QV} & J_{Q\delta} \end{pmatrix} \begin{pmatrix} \Delta|V| \\ \Delta \delta \end{pmatrix} \tag{2b}$$

Although the proposed method can be applied to various types of Distributed Generators (DGs) contributing both active and reactive power to the network, it is pertinent to note that in P2P markets, active power is predominantly traded. Furthermore, the majority of DERs are primarily designed for active power generation, exhibiting minimal reactive power generation and often operating at unity power factor. Consequently, the contribution of reactive power injection can be effectively disregarded. Therefore, the fluctuations in voltage magnitudes and voltage angles stemming from power injections and withdrawals at distinct network nodes can be succinctly characterized using equations (3a)-(3c). It is worth emphasizing that power withdrawals can be equivalently represented as negative power injections within this framework.

$$\begin{pmatrix} \Delta P \\ 0 \end{pmatrix} = \begin{pmatrix} J_{PV} & J_{P\delta} \\ J_{QV} & J_{Q\delta} \end{pmatrix} \begin{pmatrix} \Delta|V| \\ \Delta \delta \end{pmatrix} \tag{3a}$$

$$\Delta \delta = -J_{Q\delta}^{-1} J_{QV} \Delta|V|$$
$$\Delta|V| = (J_{PV} - J_{P\delta}J_{Q\delta}^{-1}J_{QV})^{-1}\Delta P \tag{3b}$$

$$S_\Delta \triangleq -J_{Q\delta}^{-1}J_{QV}S_V$$
$$S_V \triangleq (J_{PV} - J_{P\delta}J_{Q\delta}^{-1}J_{QV})^{-1} \tag{3c}$$

### B. Voltage unbalance

The location and quantity of small-scale Distributed Energy Resources (DERs) within low-voltage distribution systems are inherently stochastic processes, heavily influenced by factors such as the socio-economic status of customers and technical variables like hourly solar irradiation, particularly relevant for Photovoltaic (PV) systems. Consequently, there exists a distinct possibility that voltage unbalance may exceed the thresholds stipulated by pertinent standards. Most standards, to maintain network integrity, impose stringent limits on voltage unbalance, typically not exceeding 2%, as delineated by the Voltage Unbalance Factor (VUF), defined in equation (4) [21]. It is noteworthy that P2P transactions, particularly when involving peers from different phases, have the potential to violate this criterion, thereby accentuating the need for robust voltage unbalance management.

$$VUF = \left|\frac{V^2}{V^1}\right| \times 100 \tag{4}$$

In this paper, we introduce an analytical method aimed at incorporating power injections into the modeling of voltage unbalance. Through the application of symmetrical component analysis, we establish a connection between voltage variations in each phase and node of the network and power injections. This relationship is articulated in terms of the negative, positive, and zero sequences, as precisely defined in equations (5a)-(5c).

$$\Delta V^0 = \frac{1}{3}(\Delta V_a + \Delta V_b + \Delta V_c) \tag{5a}$$

$$\Delta V^1 = \frac{1}{3}(\Delta V_a + a\Delta V_b + a^2\Delta V_c) \tag{5b}$$

$$\Delta V^2 = \frac{1}{3}(\Delta V_a + a^2\Delta V_b + a\Delta V_c) \tag{5c}$$

The Voltage Unbalance Factor (VUF) is a crucial metric in this context, quantifying the ratio of the absolute negative sequence voltage to the positive sequence voltage at each node within the network. It is imperative that this factor remains below 2% throughout all operational periods to ensure network integrity. This requirement can be precisely formulated through a mathematical representation, expressed as a second-order cone constraint defined in equation (6). Notably, in contrast to the over/under voltage scenario, which primarily involves voltage magnitudes, modeling voltage unbalance in equation (6) necessitates the consideration of both voltage magnitudes and angles.

$$\left\| \begin{matrix} \frac{1}{3}(\Delta V_a + a^2\Delta V_b + a\Delta V_c) \\ -\frac{1}{3}\alpha(\Delta V_a + a\Delta V_b + a^2\Delta V_c) \end{matrix} \right\|_2 \leq |\alpha V^1 - V^2| \tag{6}$$

## C. congestion

The implementation of P2P energy trading within the network has the potential to alter the power flows throughout the system, potentially subjecting lines and other equipment to elevated current loads that may exceed their rated capacity. To effectively model this impact, we derive the variations in line currents by correlating them with voltage changes stemming from P2P power injections. Much like the unbalance modeling approach, the congestion model presented in this paper is formulated as a second-order cone constraint.

Equations (7a) and (7b) explicitly delineate the variations in line currents attributable to voltage fluctuations at the nodes to which these lines are connected. Notably, $\left|\Delta I_{ij}^{abc}\right|$ is the maximum allowable current limit for each line and phase.

$$\Delta I_{ij}^{abc} = \frac{\Delta V_i^{abc} - \Delta V_j^{abc}}{z_{ij}^{abc}} \tag{7a}$$

$$\left\|\Delta V_i^{abc} - \Delta V_j^{abc}\right\|_2 \le \left|z_{ij}^{abc}\right|\left|\Delta I_{ij}^{abc,\max}\right| \tag{7b}$$

## III. TRANSACTIONS EVALUATION

In this section, we present an optimization model designed to determine the acceptance or rejection of P2P transactions. The overarching objective is to maximize the inclusion of P2P trades that have been negotiated among peers and await final acceptance. This objective function operates within the bounds defined by the maximum and minimum allowable voltage magnitudes across all nodes and phases of the network, the acceptable voltage unbalance factor, and the maximum current amplitude of the lines. Formulated in equations (8a)-(8h), the problem inherently revolves around the substitution of all variables with active power injections. This approach renders it a Second Order Cone Programming (SOCP) problem, provided that all expressions within the norms are suitably modeled.

$$\begin{aligned}maximize \quad &\sum_{n \in producers} (\Delta P_n^a + \Delta P_n^b + \Delta P_n^c) \\ &- \sum_{n \in consumers} (\Delta P_n^a + \Delta P_n^b + \Delta P_n^c)\end{aligned} \tag{8a}$$

subject to:

$$\Delta V_n^{a,\min} \le \sum_n S_{V,n}^a \Delta P_n^a \le \Delta V_n^{a,\max} \qquad \forall n \tag{8b}$$

$$\Delta V_n^{b,\min} \le \sum_n S_{V,n}^b \Delta P_n^b \le \Delta V_n^{b,\max} \qquad \forall n \tag{8c}$$

$$\Delta V_n^{c,\min} \le \sum_n S_{V,n}^a \Delta P_n^c \le \Delta V_n^{c,\max} \qquad \forall n \tag{8d}$$

$$\left\|\frac{1}{3}\sum_n \begin{pmatrix} S_{V,n}^a \Delta P_n^a \angle S_{\Delta,n}^a \Delta P_n^a \\ + a^2 S_{V,n}^b \Delta P_n^b \angle S_{\Delta,n}^b \Delta P_n^b \\ + a^1 S_{V,n}^c \Delta P_n^c \angle S_{\Delta,n}^c \Delta P_n^c \end{pmatrix} - \frac{1}{3}\alpha\sum_n \begin{pmatrix} S_{V,n}^a \Delta P_n^a \angle S_{\Delta,n}^a \Delta P_n^a \\ + a^1 S_{V,n}^b \Delta P_n^b \angle S_{\Delta,n}^b \Delta P_n^b \\ + a^2 S_{V,n}^c \Delta P_n^c \angle S_{\Delta,n}^c \Delta P_n^c \end{pmatrix}\right\|_2 \le \left|\alpha V^1 - V^2\right| \quad \forall n \tag{8e}$$

$$\left\|\sum_i S_{V,i}^a \Delta P_i^a \angle S_{\Delta,i}^a \Delta P_i^a - \sum_j S_{V,j}^a \Delta P_j^a \angle S_{\Delta,j}^a \Delta P_j^a\right\|_2 \le \left|z_{ij}^a\right|\left|\Delta I_{ij}^{\max a}\right| \ \forall i,j \in la \tag{8f}$$

$$\left\|\sum_i S_{V,i}^b \Delta P_i^b \angle S_{\Delta,i}^b \Delta P_i^b - \sum_j S_{V,j}^b \Delta P_j^b \angle S_{\Delta,j}^b \Delta P_j^b\right\|_2 \le \left|z_{ij}^b\right|\left|\Delta I_{ij}^{\max b}\right| \ \forall i,j \in lb \tag{8g}$$

$$\left\|\sum_i S_{V,i}^c \Delta P_i^c \angle S_{\Delta,i}^c \Delta P_i^c - \sum_j S_{V,j}^c \Delta P_j^c \angle S_{\Delta,j}^c \Delta P_j^c\right\|_2 \le \left|z_{ij}^c\right|\left|\Delta I_{ij}^{\max c}\right| \ \forall i,j \in lc \tag{8h}$$

One approach to address this challenge is to neglect variations in voltage angles, under the assumption that the influence of power injections on voltage angles remains sufficiently small. However, in this paper, we present a linearization procedure founded on the inherent characteristics of the problem. Following a series of mathematical transformations, the details of which are elaborated in the appendix, a linear relationship between power injection and the voltage phasor is derived as:

$$\Delta V = (S_V + j\left|V\right|S_\Delta)\Delta P \times 1\angle\delta \tag{9}$$

Subsequently, by integrating this linear relationship into the optimization model, we establish a norm optimization problem. This problem can be further transformed into a Second Order Cone Problem (SOCP) amenable to efficient algorithmic solutions. The final SOCP model, representing the evaluation process for P2P energy trading, is formulated as equations (10a)-(10h). This analytical model can be effectively solved using efficient algorithms, including interior point methods [22, 23].

$$\begin{aligned}maximize \quad &\sum_{n \in producers} (\Delta P_n^a + \Delta P_n^b + \Delta P_n^c) \\ &- \sum_{n \in consumers} (\Delta P_n^a + \Delta P_n^b + \Delta P_n^c)\end{aligned} \tag{10a}$$

subject to:

$$\Delta V_n^{a,\min} \le \sum_n S_{V,n}^a \Delta P_n^a \le \Delta V_n^{a,\max} \qquad \forall n \tag{10b}$$

$$\Delta V_n^{b,\min} \le \sum_n S_{V,n}^b \Delta P_n^b \le \Delta V_n^{b,\max} \qquad \forall n \tag{10c}$$

$$\Delta V_n^{c,\min} \le \sum_n S_{V,n}^a \Delta P_n^c \le \Delta V_n^{c,\max} \qquad \forall n \tag{10d}$$

$$\left\|\frac{1}{3}\sum_n \begin{pmatrix} ((S_{V,n}^a + j\left|V_n^a\right|S_{\Delta,n}^a)\Delta P_n^a \times 1\angle\delta_n^a \\ a^2(S_{V,n}^b + j\left|V_n^b\right|S_{\Delta,n}^b)\Delta P_n^b \times 1\angle\delta_n^b \\ a^1(S_{V,n}^c + j\left|V_n^c\right|S_{\Delta,n}^c)\Delta P_n^c \times 1\angle\delta_n^c) \end{pmatrix} - \frac{1}{3}\alpha\sum_n \begin{pmatrix} ((S_{V,n}^a + j\left|V_n^a\right|S_{\Delta,n}^a)\Delta P_n^a \times 1\angle\delta_n^a \\ a^1(S_{V,n}^b + j\left|V_n^b\right|S_{\Delta,n}^b)\Delta P_n^b \times 1\angle\delta_n^b \\ a^2(S_{V,n}^c + j\left|V_n^c\right|S_{\Delta,n}^c)\Delta P_n^c \times 1\angle\delta_n^c) \end{pmatrix}\right\|_2 \le \left|\alpha V^1 - V^2\right| \quad \forall n \tag{10e}$$

$$\left\|\begin{array}{l}\sum_i (S^a_{V,i} + j\left|V^a_i\right|S^a_{\Delta,i})\Delta P^a_i \times 1\angle\delta^a_i \\ -\sum_j (S^a_{V,j} + j\left|V^a_j\right|S^a_{\Delta,j})\Delta P^a_j \times 1\angle\delta^a_j\end{array}\right\|_2 \leq \left|z^a_{ij}\right|\left|\Delta I^{\max a}_{ij}\right| \quad \forall i,j \in la$$

$$(10f)$$

$$\left\|\begin{array}{l}\sum_i (S^b_{V,i} + j\left|V^b_i\right|S^b_{\Delta,i})\Delta P^b_i \times 1\angle\delta^b_i \\ -\sum_j (S^b_{V,j} + j\left|V^b_j\right|S^b_{\Delta,j})\Delta P^b_j \times 1\angle\delta^b_j\end{array}\right\|_2 \leq \left|z^b_{ij}\right|\left|\Delta I^{\max b}_{ij}\right| \quad \forall i,j \in lb$$

$$(10g)$$

$$\left\|\begin{array}{l}\sum_i (S^c_{V,i} + j\left|V^c_i\right|S^c_{\Delta,i})\Delta P^c_i \times 1\angle\delta^c_i \\ -\sum_j (S^c_{V,j} + j\left|V^c_j\right|S^c_{\Delta,j})\Delta P^c_j \times 1\angle\delta^c_j\end{array}\right\|_2 \leq \left|z^c_{ij}\right|\left|\Delta I^{\max c}_{ij}\right| \quad \forall i,j \in lc$$

$$(10h)$$

## IV. MARKET MECHANISM

P2P energy markets exhibit distinctive characteristics that set them apart from traditional wholesale and local energy markets. In stark contrast to wholesale markets, P2P markets do not necessitate a global balancing of total supply and demand. This implies that while participants must be suitably motivated to share their assets within the market, the responsibility for meeting the entire demand can rest with the Distribution System Operator (DSO) or other entities operating under various mechanisms.

Another significant departure from conventional wholesale markets is that the DSO or the entity administering the P2P energy platform is not permitted to accept or reject transactions on economic grounds. Unlike wholesale markets, where supply-demand curves and models determine the selection of generators and the allocation of load, P2P energy markets are characterized by a high degree of democratization. Here, only peers have the authority to accept or decline economic terms associated with a transaction. However, the DSO retains the discretion to approve or deny transactions negotiated in terms of price and energy quantity based on network constraints.

Furthermore, participants in P2P energy markets are expected to undergo long-term training to develop strategies aimed at maximizing their profits. This necessitates the coordination of local markets with P2P markets to ensure the rational behavior of the market ecosystem.

Another pivotal characteristic that sets P2P energy markets apart from bilateral markets is their distinct nature. In contrast to bilateral contracts, which primarily serve as risk management instruments for suppliers and consumers, P2P energy markets are not designed for risk mitigation. P2P markets are characterized by frequent openings and closings throughout the day, sometimes occurring within the span of an hour. In contrast, bilateral contracts typically involve long-term commitments and substantial quantities of energy.

Furthermore, the finalization of a bilateral contract is a time-consuming and costly process, often necessitating third-party involvement to guarantee the contract terms. This can result in added expenses for both parties. Conversely, P2P transactions involve significantly smaller quantities of energy and are characterized by their short-term, cost-effective nature.

Drawing upon the distinctive features of P2P energy markets, we propose the Block Double Auction (BDA) mechanism as a well-suited market scheme, structured as follows:

- Sellers participate by offering their surplus energy along with corresponding prices for the designated time slot T, drawing upon their demand and generation forecasts.
- Buyers enter the market by presenting their demand requirements and the prices they are willing to pay for the specified time slot T, rooted in their demand forecasts.
- The Double Auction mechanism orchestrates the matching of sellers and buyers based on their respective bids and offers.
- The resulting block of matched pairs is then submitted to the Distribution System Operator (DSO) for evaluation. The DSO assesses these transactions against network constraints, guided by the base case conditions observed in the preceding time slot T-1.
- Following the DSO's review, the market session concludes.
- Subsequently, the entire process recommences for time slot T+1, as participants engage in the next trading period.

In the Block Double Auction (BDA) mechanism, we depart from the conventional approach of evaluating transactions individually. Such a method proves inadequate in scenarios where each transaction's impact on network constraints may influence other transactions. Our BDA mechanism offers a holistic assessment of a block of transactions, recognizing that the interactions among transactions can significantly affect network conditions.

Notably, this mechanism facilitates transactions between peers, even when they are not on the same phase within the distribution system. The collective block of transactions, as evaluated by the DSO, possesses the potential to rectify voltage unbalance, over/under voltage, and congestion issues that may arise from individual transactions.

Finally, equations (11a)-(11e) have been incorporated into the P2P evaluation model presented in equation (10). These additions serve to effectively model the distinct characteristics that define a P2P energy market.

$$\Delta P_{n\in producers} + \Delta P_{m\in consumers} = 0 \quad \forall (m,n) \tag{11a}$$

$$\Delta P_{n\in producers} \geq 0 \quad \forall n \in producers \tag{11b}$$

$$\Delta P_{m\in consumers} \leq 0 \quad \forall m \in consumers \tag{11c}$$

$$\Delta P_m = 0 \quad \forall m \notin consumers \tag{11d}$$

$$\Delta P_n = 0 \quad \forall n \notin producers \tag{11e}$$

In the realm of P2P energy markets, characterized by their short-term and small-scale transactions, it becomes imperative for peers to have clarity regarding transaction fees. A significant portion of these fees can be attributed to technical losses within the network. Furthermore, evaluating the impact of each individual transaction on these losses proves challenging, as it is the aggregate of transactions that ultimately determines whether a transaction contributes to a reduction in system losses. Consequently, the proposed Block Double Auction (BDA) method for the market mechanism offers an advantage when it comes to assessing transaction fees.

In light of these considerations, this paper introduces the concept of the Transaction Loss Coefficient (TLC) as a means of quantifying the impact of each transaction on losses within the network. (12a) is obtained as the coefficient of injection at node k ($\Delta P_k$) of the line loss which nodes i and j are connected to ($\Delta P_{i,j}^{loss}$). The mathematical procedure to get (12a) is provided in the appendix.

$$\frac{\partial \Delta P_{i,j}^{loss}}{\partial \Delta P_k} = \frac{2R_{i,j}}{\left|z_{i,j}\right|^2} (\text{Re}(A)\text{Re}(C) + \text{Im}(A)\text{Im}(C) + \text{Re}(B)\text{Re}(D)$$
$$+ \text{Im}(B)\text{Im}(D) - \text{Re}(B)\text{Im}(C) + \text{Re}(C)\text{Im}(B)$$
$$- \text{Re}(A)\text{Im}(D) + \text{Re}(D)\text{Im}(A))$$
(12a)

$$A = S_V^{i,k}\angle\delta_i - S_V^{j,k}\angle\delta_j$$
(12b)

$$B = \left|V_i\right|S_\Delta^{i,k}\angle\delta_i - \left|V_j\right|S_\Delta^{j,k}\angle\delta_j$$
(12c)

$$C = \sum_n (S_V^{i,n}\angle\delta_i - S_V^{j,n}\angle\delta_j)\Delta P_n$$
(12d)

$$D = \sum_n (\left|V_i\right|S_\Delta^{i,n}\angle\delta_i - \left|V_j\right|S_\Delta^{j,n}\angle\delta_j)\Delta P_n$$
(12e)

It is important to note that each transaction comprises two injections: one positive and the other negative. Accordingly, the Transaction Loss Coefficient (TLC) for energy trading between peer m and peer n is defined as follows:

$$TLC_{m,n} = \sum_l (\frac{\partial \Delta P_l^{loss}}{\partial \Delta P_m}\Delta P_m + \frac{\partial \Delta P_l^{loss}}{\partial \Delta P_n}\Delta P_n)$$
(13)

## V. NUMERICAL STUDY

this section, we introduce a modified IEEE 33-bus distribution test system that has been tailored for simulation and numerical study purposes. The modification primarily involves adjusting the voltage levels from 12.6 kV to 0.4 kV and modifying the loads of the system as detailed in Table I. All other parameters of the system remain consistent with the standard test configuration. For reference, the network topology is illustrated in Figure 1 [24].

It's important to note that in this setup, every node on each phase is assumed to function as a prosumer, such as Microgrids (MGs), with the exception of node 1, which serves as the substation node connecting the system to the upstream network.

To adhere to technical standards and research findings, we have established certain parameters for this study. The maximum allowable voltage deviation at each node is set at 0.05 per unit (P.U.), while the maximum permissible voltage unbalance factor is capped at 2%, in line with industry standards and research literature. Additionally, it is presumed that up to 30% of the base case current can be incrementally added to each line's current flow.

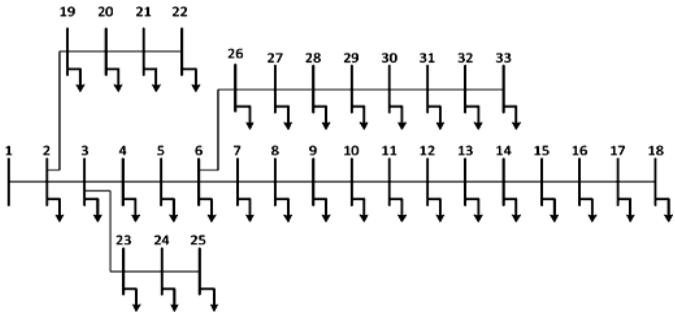

*Fig.1. The IEEE 33-bus radial distribution system*

| Bus | P(KW) | Q(KVAR) | Bus | P(KW) | Q(KVAR) |
|-----|-------|---------|-----|-------|---------|
| 1 | 0 | 0 | 17 | 3 | 1 |
| 2 | 5 | 3 | 18 | 4.5 | 2 |
| 3 | 4.5 | 2 | 19 | 4.5 | 2 |
| 4 | 6 | 4 | 20 | 4.5 | 2 |
| 5 | 3 | 1.5 | 21 | 4.5 | 2 |
| 6 | 3 | 1 | 22 | 4.5 | 2 |
| 7 | 10 | 5 | 23 | 4.5 | 2.5 |
| 8 | 10 | 5 | 24 | 21 | 10 |
| 9 | 3 | 1 | 25 | 21 | 10 |
| 10 | 3 | 1 | 26 | 3 | 1.25 |
| 11 | 2.25 | 1.5 | 27 | 3 | 1.25 |
| 12 | 3 | 1.75 | 28 | 3 | 1 |
| 13 | 3 | 1.75 | 30 | 6 | 3.5 |
| 14 | 6 | 4 | 31 | 10 | 30 |
| 15 | 3 | 0.5 | 32 | 7.5 | 3.5 |
| 16 | 3 | 1 | 33 | 10.5 | 5 |

### A. P2P transactions on the same phase

In this particular scenario, the assumption is that prosumers within each phase can exclusively engage in energy trading with their counterparts on the same phase. Specifically, we consider a scenario where producers (Pa = {5, 7, 15, 19}) aim to sell a specified quantity 'X' of energy during time slot T to consumers (Ca = {24, 17, 3, 32}). Importantly, no transactions are to be conducted on phases B and C. The objective here is to ascertain the maximum power trade that can be facilitated. The acceptance of a transaction is contingent upon the proposed model's determination that the maximum allowable power trades, as determined by the model, exceed the negotiated power quantity agreed upon by the peers.

To initiate the proposed model, it is imperative to first solve the base case. In this context, we assume that the loads on phase A are consistent with Table 1. Concurrently, loads on phase B are 20% higher than those on phase A, while loads on phase C are 20% lower than those on phase A. Figure 2 provides a visual representation of the voltage magnitudes at nodes for each phase within the base case.

Table II provides the results of the process of P2P transactions evaluation for different limit bounds of the criteria. As can be seen, in the given operational condition, the congestion limit has the most impact on the allowable amount of power that the peers can trade. Voltage deviation and voltage unbalance come next, respectively. For example, when $\alpha = 1$, $\Delta I_{\max} = 30$ and $\Delta V_{\max} = 5$, the maximum power to trade between the given peers is 132, 51, 110 and 98 Kw while for $\Delta I_{\max} = 10$ and the same value for the other criteria the Peers can trade up to 44, 17, 36, and 32 Kw, respectively.

It is interesting to know that if the transactions to be evaluated individually, the maximum power for trade without violating the network constraint are 132, 51, 59 and 98 Kw for the peers, respectively, when $\alpha = 1$, $\Delta I_{\max} = 30$ and $\Delta V_{\max} = 5$. When $\Delta I_{\max}$ is set to be 10 and the other parameters are the same as before, the maximum allowable power to trade for peers {15,3} drops to 19 Kw while the other transactions maximum power remains at the same level when collectively assessed. Note that, it is evident that there may be the case in which the maximum power is higher when the transactions are individually evaluated and vice versa. The point is that all the transactions must be collectively evaluated as the way the proposed BDA market mechanism offers.

Figures 3-5 depicts the maximum allowable P2P transactions on phases A, B and C, respectively, when producer peers and consumer peers are determined randomly. In such a given situation, the total P2P transaction capacity trade for each phase is 1128, 1056 and 1135 Kw, respectively. Fig.6 shows the maximum P2P power each transaction can trade, when they are sorted from the lowest to the highest. For example, roughly the half of the transactions cannot have the power more than 100 Kw.

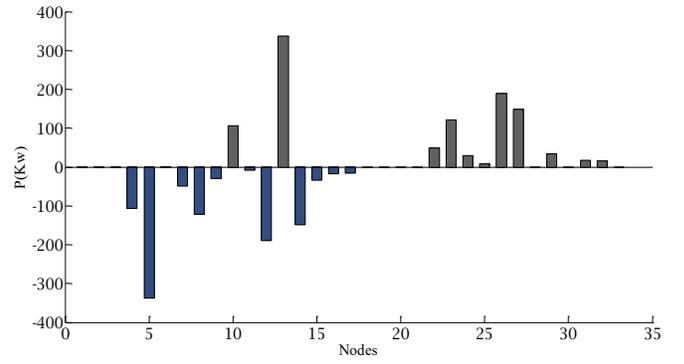

*Fig.4. The maximum allowable transactions for phase B*

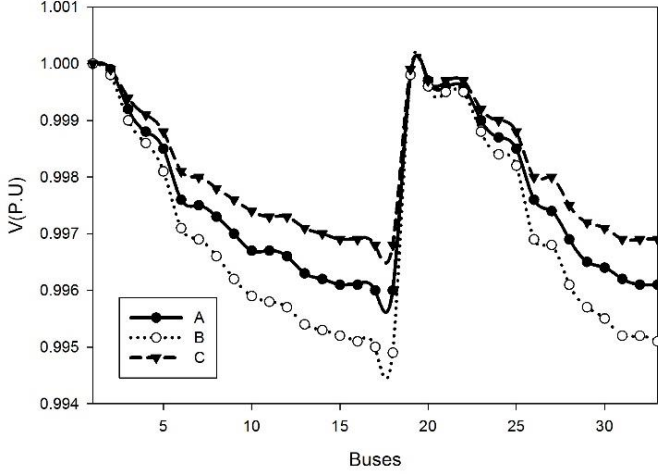

*Fig.2. The voltage magnitudes of buses in case base*

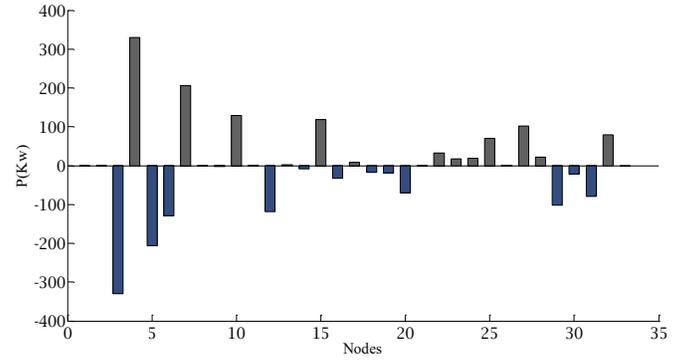

*Fig.5. The maximum allowable transactions for phase C*

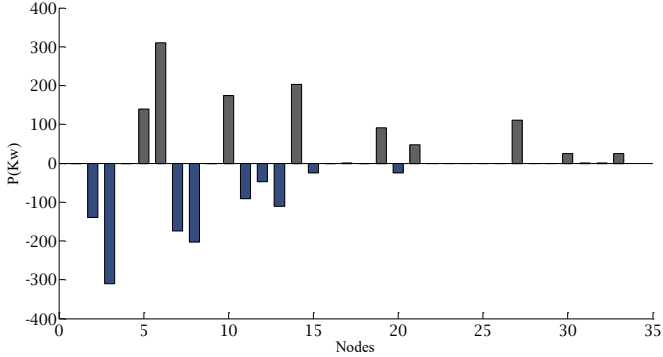

*Fig.3. The maximum allowable transactions for phase A*

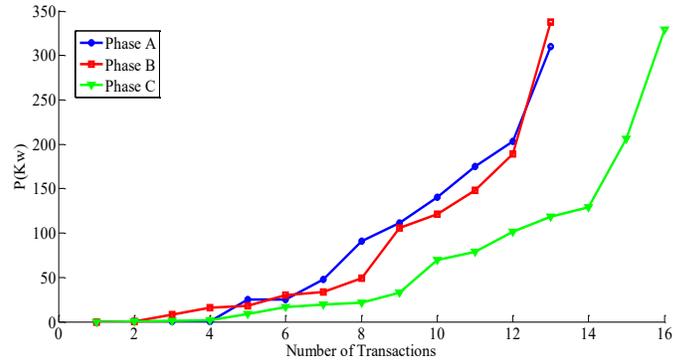

*Fig.6. The maximum allowable trade capacity for each transaction from the lowest to the highest*

TABLE II: THE MAXIMUM ALLOWABLE P2P TRANSACTIONS WITH DIFFERENT CRITERIA

| Producers | Consumers | $P^{\max}(Kw)$ | $\alpha(\%)$ | $\Delta I_{\max}(\%)$ | $\Delta V_{\max}(\%)$ |
|---|---|---|---|---|---|
| {5,7,15,19} | {24,17,3,32} | {132, 51, 110, 98} | 1 | 30 | 5 |
| {5,7,15,19} | {24,17,3,32} | {132, 51, 110, 98} | 2 | 30 | 5 |
| {5,7,15,19} | {24,17,3,32} | {44, 17, 36, 32} | 1 | 10 | 5 |
| {5,7,15,19} | {24,17,3,32} | {221, 85, 184, 163} | 1 | 50 | 5 |
| {5,7,15,19} | {24,17,3,32} | {221, 85, 184, 163} | 1 | 50 | 10 |
| {5,7,15,19} | {24,17,3,32} | {884, 341, 736, 655} | 1 | 200 | 5 |
| {5,7,15,19} | {24,17,3,32} | {884, 342, 574, 533} | 1 | 200 | 10 |
| {5,7,15,19} | {24,17,3,32} | {2211, 855, 1239, 1150} | 1 | 500 | 5 |
| {5,7,15,19} | {24,17,3,32} | {2221, 854, 1841, 1638} | 2 | 500 | 5 |

## B. P2P transactions between different phases

While single-phase trades often prevail in P2P energy markets, they can introduce challenges such as voltage unbalance, alongside opportunities like the concept of a voltage balance service provider. It's important to note that these issues, while intriguing, fall beyond the immediate scope of this paper and warrant consideration as potential subjects for future research.

The primary objective of this paper, however, centers on the assessment of P2P transactions, particularly when a significant portion or even the entirety of these transactions take place across different phases. In this specific scenario, we consider a scenario where producer peers (Pa = {5, 7, 15, 19}) on phase A have engaged in negotiations with consumer peers (Cb = {24, 17, 3, 32}) on phase B, with the goal of delivering a specified quantity of energy. The proposed method is then employed to evaluate the feasibility of these transactions, with the base case remaining consistent with the description provided earlier.

Table III presents the outcomes pertaining to the maximum allowable power capacity for P2P trade. The results yield several intriguing insights. Firstly, there is a noticeable decrease in the overall P2P transaction capacity when compared to Table II, where the trades occurred within the same phase. This shift underlines the impact of cross-phase transactions on the total trade capacity.

Secondly, a significant revelation is that while the aggregate capacity of P2P transactions is predominantly influenced by the congestion criterion, individual P2P transactions are profoundly affected by the voltage unbalance criterion. These findings underscore the nuanced dynamics at play within P2P energy markets, where a delicate interplay of factors shapes the trade landscape. For instance, when $\Delta I_{max} = 30$ and $\Delta V_{max} = 5$, the total capacity is 317 Kw for $\alpha = 1$ and $\alpha = 2$ whereas the maximum allowable power of each transaction is different. The same thing can be observed for $\alpha = 2$ and $\alpha = 5$ when $\Delta I_{max} = 200$ and $\Delta V_{max} = 5$. The third significant point is that there can be sort of condition depending on the base case operational condition and the matched peers in the P2P energy market, like this one, in which if a transaction takes place with either a certain amount of power or more than that, another transaction will not be allowed to happen at all. For instance, in our case, the maximum allowable transaction power between peers belonging to the nodes 7 to 17 is zero when $\Delta I_{max} = 200$, $\Delta V_{max} = 5$ or $\Delta V_{max} = 10$ and $\alpha = 1$.

However, in an individual transaction assessment, it becomes apparent that a transaction could be approved with a power capacity of up to 313 kW. It's crucial to note that the entity responsible for deciding whether to accept or reject a transaction cannot approve it as long as the other transactions have settled on a specified level of power capacity that aligns with the evaluation process. Unless, of course, the negotiations for the other transactions are structured in a manner that results in power capacities below the maximum allowable level.

Fig. 7 provides a clear visualization of this concept. It illustrates how the maximum allowable power for each transaction varies based on the negotiated power trades between peers. For instance, transaction 7-17 can only be accepted when the other three transactions have negotiated power quantities that are collectively less than 340 kW. This pattern holds true for all transactions; however, in this specific case, transaction 5-24 emerges as dominant and more capable in terms of power capacity. Consequently, in such scenarios, transactions 5-24, 19-32, and 15-3 are technically better matched.

## C. Transactions loss allocation

This section delves into a critical component of the settlement procedure within P2P markets—namely, the allocation of loss costs for each transaction and, consequently, for each peer. Leveraging the methodology proposed in the market mechanism, we conduct a study in this section to evaluate the effectiveness of this approach for loss allocation.

For the purposes of this analysis, we consider the first-row transactions from Tables II and III, both conducted with the maximum allowable powers. Table IV offers insights into the Transaction Loss Coefficients (TLCs) for producer and consumer peers when transactions occur exclusively within phase A. Notably, the transaction between nodes 15 and 3 exhibits the highest total TLC among the four transactions, while the transaction between nodes 7 and 17 demonstrates a negative total TLC. This negative value suggests that this particular transaction contributes to the reduction of network losses.

Similarly, Table V provides an overview of the TLCs for transactions negotiated and executed between different phases, again with the maximum allowable powers. It's important to observe that, in such scenarios, all transactions exhibit higher TLC values compared to when transactions occurred exclusively within phase A. Notably, the TLC for the transaction 15-3 decreases due to the reduction in maximum allowable power.

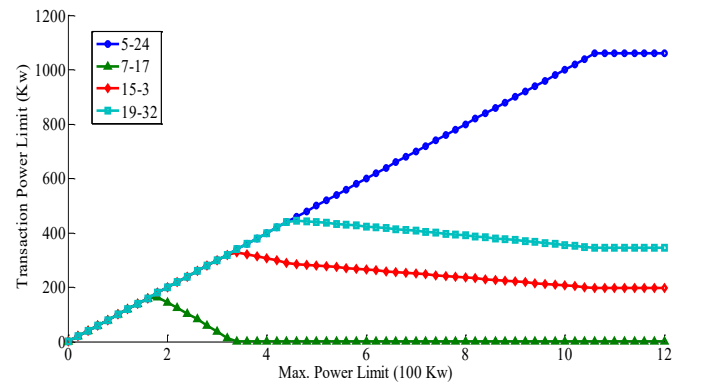

*Fig.7. The maximum allowable trade capacity for each transaction with regard to each transaction negotiated power*

TABLE IV: THE PEER SHARES IN EACH TRANSACTION LOSS ON PHASE A

| Producer | Consumer | $\{TLC_m(Kw), TLC_n(Kw)\}$ |
|----------|----------|------------------------------|
| {5} | {24} | {0.0965, 0.3768} |
| {7} | {17} | {0.0559, -0.2164} |
| {15} | {3} | {0.6074, 0.0671} |
| {19} | {32} | {0.0197, 0.4162} |

TABLE III: THE MAXIMUM ALLOWABLE P2P TRANSACTIONS BETWEEN PHASES A AND B WITH DIFFERENT CRITERIA

| Producers (A) | Consumers (B) | $P^{\max}(Kw)$ | $\alpha(\%)$ | $\Delta I_{\max}(\%)$ | $\Delta V_{\max}(\%)$ |
|---|---|---|---|---|---|
| {5,7,15,19} | {24,17,3,32} | {138, 40, 38, 101} | 1 | 30 | 5 |
| {5,7,15,19} | {24,17,3,32} | {148, 41, 35, 93} | 2 | 30 | 5 |
| {5,7,15,19} | {24,17,3,32} | {50, 18, 9, 28} | 1 | 10 | 5 |
| {5,7,15,19} | {24,17,3,32} | {216, 70, 88, 156} | 1 | 50 | 5 |
| {5,7,15,19} | {24,17,3,32} | {222, 72, 74, 162} | 1 | 50 | 10 |
| {5,7,15,19} | {24,17,3,32} | {1061, 0, 198, 344} | 1 | 200 | 5 |
| {5,7,15,19} | {24,17,3,32} | {1061, 0, 198, 344} | 1 | 200 | 10 |
| {5,7,15,19} | {24,17,3,32} | {848, 213, 338, 726} | 2 | 200 | 5 |
| {5,7,15,19} | {24,17,3,32} | {949, 339, 230, 607} | 5 | 200 | 5 |
| {5,7,15,19} | {24,17,3,32} | {19, 3, 11, 7} | 0 | 200 | 5 |
| {5,7,15,19} | {24,17,3,32} | {878, 376, 223, 648} | - | 200 | 5 |

TABLE V: THE PEER SHARES IN EACH TRANSACTION LOSS ON PHASES A AND B

| Producer (A) | Consumer (B) | $\{TLC_m(Kw), TLC_n(Kw)\}$ |
|---|---|---|
| {5} | {24} | {0.5140, 0.6455} |
| {7} | {17} | {0.1918, 0.3782} |
| {15} | {3} | {0.2946, 0.0897} |
| {19} | {32} | {0.0582, 1.0354} |

## VI. CONCLUSION

In this paper, a comprehensive framework has been presented for Peer-to-Peer (P2P) energy trading. This approach addresses both significant technical challenges and market mechanisms, culminating in the development of a novel method to assess transaction viability within the technical constraints of the network. We have rigorously considered over/under voltage, voltage unbalance, and congestion, devising a second-order cone programming model that empowers the Distribution System Operator (DSO) to make reliable decisions regarding the acceptance or rejection of P2P transactions.

From the perspective of market mechanisms, the concept of a block double auction has been introduced, recognizing that each transaction's behavior has a ripple effect on others. The technical constraints that prohibit the application of a standard double auction procedure within P2P electricity markets have been underscored. Furthermore, a method for loss allocation within P2P markets have been devised, leveraging insights gleaned from our proposed transaction evaluation model.

This methodology underwent rigorous testing using the IEEE 33-bus test system, demonstrating the prowess of our framework in addressing the unique challenges posed by P2P markets. Notably, the maximum allowable power capacity was determined through peer negotiations. Peers can engage in trades up to the maximum allowable power, a value contingent upon the operational conditions and network constraint criteria.

The findings revealed that when peers originate from the same phase, the total power capacity available for trade surpasses that when peers hail from different phases. Moreover, it was evident that the dominant network constraint is congestion or voltage violation (depending on the operational conditions) when peers trade within the same phase. Conversely, voltage unbalance emerges as the predominant constraint when peers originate from different phases.

Another noteworthy conclusion is that the maximum allowable power capacity for a transaction is intrinsically tied to the allowable capacity consumed by other transactions. Additionally, we observed that system losses are higher when transactions involve peers from different phases compared to transactions confined to the same phase.

## APPENDIX

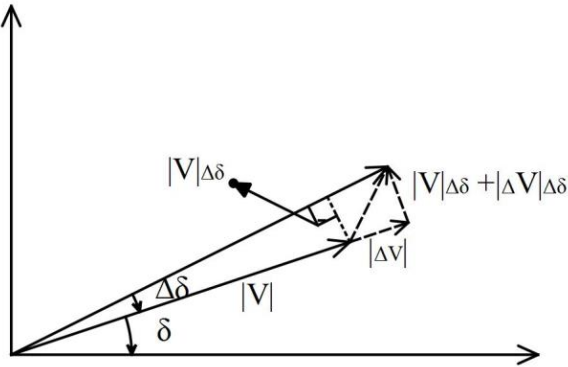

### A. linearization

$$V^{new} = V^{old} + \Delta V = |V| \angle \delta + \Delta V$$

$$(|V| + |\Delta V|)\angle(\delta + \Delta\delta) = (|V| + |\Delta V|)(\cos(\delta + \Delta\delta) + j\sin(\delta + \Delta\delta))$$

$$=(|V| + |\Delta V|)(\cos(\delta)\cos(\Delta\delta) - \sin(\delta)\sin(\Delta\delta) + j\sin(\delta)\cos(\Delta\delta) + j\sin(\Delta\delta)\cos(\delta))$$

$$\sin\Delta\delta = \Delta\delta$$

$$\cos\Delta\delta = 1$$

$$\Delta V = (\cos(\delta) + j\sin(\delta))(|\Delta V| + j|V|\Delta\delta + j|\Delta V|\Delta\delta)$$

$$|\Delta V|\Delta\delta \approx 0$$

$$\Delta V = (\cos(\delta) + j\sin(\delta))(|\Delta V| + j|V|\Delta\delta)$$

$$\Delta V = (\cos(\delta) + j\sin(\delta))(S_V\Delta P + j|V|S_\Delta\Delta P)$$

$$\Delta V = (S_V + j|V|S_\Delta)\Delta P \times 1\angle\delta$$

### B. transaction loss coefficient

$$\Delta P_{i,j}^{loss} = R_{i,j}|\Delta I_{i,j}|^2 = R_{i,j}\Delta I_{i,j}^*\Delta I_{i,j}$$

$$\frac{\partial\Delta P_{i,j}^{loss}}{\partial\Delta P_k} = R_{i,j}(\frac{\partial\Delta I_{i,j}^*}{\partial\Delta P_k}\Delta I_{i,j} + \frac{\partial\Delta I_{i,j}}{\partial\Delta P_k}\Delta I_{i,j}^*)$$

$$\frac{\partial\Delta I_{i,j}}{\partial\Delta P_k} = \frac{(S_V^{i,k} + j|V_i|S_\Delta^{i,k})\times 1\angle\delta_i - (S_V^{j,k} + j|V_j|S_\Delta^{j,k})\times 1\angle\delta_j}{z_{i,j}}$$

$$= \frac{(S_V^{i,k}\angle\delta_i - S_V^{j,k}\angle\delta_j) + j(|V_i|S_\Delta^{i,k}\angle\delta_i - |V_j|S_\Delta^{j,k}\angle\delta_j)}{z_{i,j}}$$

$$\Delta I_{i,j} = \frac{\sum_n(S_V^{i,n} + j|V_i|S_\Delta^{i,n})\Delta P_n\times 1\angle\delta_i - \sum_n(S_V^{j,n} + j|V_j|S_\Delta^{j,n})\Delta P_n\times 1\angle\delta_j}{z_{i,j}}$$

$$= \frac{\sum_n((S_V^{i,n}\angle\delta_i - S_V^{j,n}\angle\delta_j) + j(|V_i|S_\Delta^{i,n}\angle\delta_i - |V_j|S_\Delta^{j,n}\angle\delta_j))\Delta P_n}{z_{i,j}}$$

$$A = S_V^{i,k}\angle\delta_i - S_V^{j,k}\angle\delta_j$$

$$B = |V_i|S_\Delta^{i,k}\angle\delta_i - |V_j|S_\Delta^{j,k}\angle\delta_j$$

$$C = \sum_n(S_V^{i,n}\angle\delta_i - S_V^{j,n}\angle\delta_j)\Delta P_n$$

$$D = \sum_n^n(|V_i|S_\Delta^{i,n}\angle\delta_i - |V_j|S_\Delta^{j,n}\angle\delta_j)\Delta P_n$$

$$\frac{\partial\Delta P_{i,j}^{loss}}{\partial\Delta P_k} = \frac{2R_{i,j}}{|z_{i,j}|^2}(\text{Re}(A)\,\text{Re}(C) + \text{Im}(A)\,\text{Im}(C) + \text{Re}(B)\,\text{Re}(D) + \text{Im}(B)\,\text{Im}(D) - \text{Re}(B)\,\text{Im}(C) + \text{Re}(C)\,\text{Im}(B) - \text{Re}(A)\,\text{Im}(D) + \text{Re}(D)\,\text{Im}(A))$$